# Practical development of efficient thermoelectric – photovoltaic hybrid systems based on wide-gap solar cells

Bruno Lorenzi,*[1] Paolo Mariani[2], Andrea Reale[2], Aldo Di Carlo[2,3], Gang Chen[4], Dario Narducci[1]

[1]Dept. Materials Science, Univ. of Milano Bicocca, Milano, 20125, Italy
[2]CHOSE, Center for Hybrid and Organic Solar Energy, University of Rome Tor Vergata, Rome, 00133 Italy
[3]ISM-CNR, Institute of Structure of Matter, National Research Council, Rome, 00133 Italy
[3]Dept. Mechanical Engineering, Massachusetts Institute of Technology, Cambridge, MA, 02139, USA

The decrease of solar cell efficiency with temperature is a known problem for photovoltaics (PV). Temperature sensitivity can lead to a considerable amount of energy losses over the lifetime of solar panels. In this perspective Hybrid Thermoelectric-Photovoltaic (HTEPV) systems, which recover solar cell heat losses to produce an additional power output, can be a suitable option. However only hybridization of wide-gap solar cells is convenient in terms of efficiency gains and deserves investigation to evaluate HTEPV devices effectiveness. In this work we report the modeling and the development of customized bismuth telluride thermoelectric generators, optimized to be hybridized with amorphous silicon (aSi), Gallium Indium Phosphide (GaInP) or Perovskites solar cells. The model results showed in all three cases efficiency gains with a maximum of +3.1% for Perovskites (from 16.4% to 19.5%). These enhancements were then experimentally validated for the case of Perovskites solar cells, for which maximum gains were found to occur at typical operating temperatures of conventional PVs. This experimental evaluation demonstrated in an accurate fashion the real potential of thermoelectric hybridization of solar cells.

*corresponding author: bruno.lorenzi@unimib.it

## 1 Introduction

Silicon solar cells dominate the photovoltaic (PV) market. Their high efficiencies, along with their relatively low cost, make them the most valuable solution nowadays. However, it is well known that silicon solar cells are extremely sensitive to temperature, and that they can lose up to 15 – 20% of their room temperature efficiency under normal operating conditions [1,2]. Wide-gap solar cells have been proposed as valuable alternative solutions, because to their smaller temperature sensitivity[3]. However, the room temperature efficiencies of these candidate materials are still lower than silicon-based devices, making them not competitive enough. The reasons for these lower performances is a smaller readiness level of the materials and of the devices based on wide-gap absorbers [4].

A suitable option to enhance the competitiveness of wide-gap solar cells more rapidly is to implement heat recovery strategies. It is well know that for single-junction solar cells the efficiency are limited by the so-called Shockley-Quisser limit [5], and that most of the losses in solar cells (up to 90%) are in the form of heat [6–9]. Among all the suitable options in order to enhance PV efficiencies by heat recovery, the hybridization with thermoelectric generators (TEGs), has gained an increasing attention in recent years. In Hybrid Thermoelectric-Photovoltaic (HTEPV) systems, the TEG part can recover the solar cell heat to produce an additional power output and enhance the overall device output power and efficiency. HTEPV devices have been the object of many studies and reviews, especially in the last years [10–12]. Within this literature the opinion on the actual effectiveness of HTEPV devices has been quite fluctuating. Actually along the years these systems have been reported as both very convenient [13–16] and not suitable to increase PV efficiency [17,18]. In the last couple of years most of the works published on HTEPV systems have been dedicated to the study of their transient behavior over time [19–21], to the effect of the TEG geometrical parameters [22,23], and to hybrids with perovskites solar cells [24,25]. In particular for perovskites, Zhou et al. [24] reported a surprising 4.9% efficiency gain for a $CH_3NH_3Pb(I_{0.95}Br_{0.05})_3$ solar cell with optimized energy gap, coupled with a commercial TEG. Strangely enough the authors report a ΔT over the TEG sides of only 7.7 °C corresponding to the mentioned efficiency gain, making the interpretation of this result quite difficult. A possible explanation can be related to the different renormalization of the power densities over the respective areas of the two parts of the hybrid device. Fu et al. [25] reported instead the hybridization of a large area perovskite solar module with a customized TEG based on bismuth telluride. In this case, the authors showed a more sensible 1% efficiency gain reaching a total hybrid efficiency of 12.7%. The work is particularly interesting for the study of the electrical hybridization of the two system parts, and for the analysis of the efficiency stability over time.

These preliminary results on the thermoelectric hybridization with perovskite solar cells, call for further analysis of the potential of HTEPV systems with this kind of material, and in general with wide-gap solar cells. This is the approach we follow in this work.

In fact here we present a theoretical and experimental study on HTEPV systems based on wide-gap solar cells. Based on a detailed characterization of three kinds of wide-gap cells, namely amorphous silicon (aSi), Gallium Indium Phosphide (GaInP) and Perovskites solar cells, we exploited a theoretical model to predict the hybrid system behavior versus temperature. The model results were then used to choose the best PV material for the thermoelectric hybridization (namely the perovskite solar cell), and to define the optimal TEG design. An optimized bismuth telluride TEG was then developed and characterized on the basis of previous studies on pure Solar Thermoelectric Generators (STEGs) [26–28]. Finally, the HTEPV devices were built and characterized as a function of temperature, optical concentration, ambient pressure, and cold side cooling. This experimental evaluation validated the theoretical model used in this work and demonstrated the actual potential of HTEPV system based on wide-gap solar cells.

## 2 System description and modeling

The hybrid system developed in this work consists of an optimized TEG placed in thermal contact with the back of a wide-gap solar cells, as depicted in Fig. 1. This kind of thermally coupled generators should not be confused with another class of HTEPV systems, normally referred as optically coupled devices[29,30]. In that case the TEG and PV parts are in fact separated, and a spectrum splitting technology is implemented to direct different parts of the solar spectrum towards them. In a previous study we showed how the optically coupled approach is less convenient than the thermally coupled one [31]. When the TEG is not in thermal contact with the solar cell, it can only access the infrared (IR) part of the solar spectrum, without being able to recover heat losses of the PV part. Most of the heat losses are in fact happening in the UV-Vis section of the spectrum where the solar cells absorb [9]. In other words, it is more convenient, in terms of final efficiency gains, to keep the solar cell at the same temperature of the TEG hot side, instead of keeping the cell cold but loosing much of the recoverable heat.

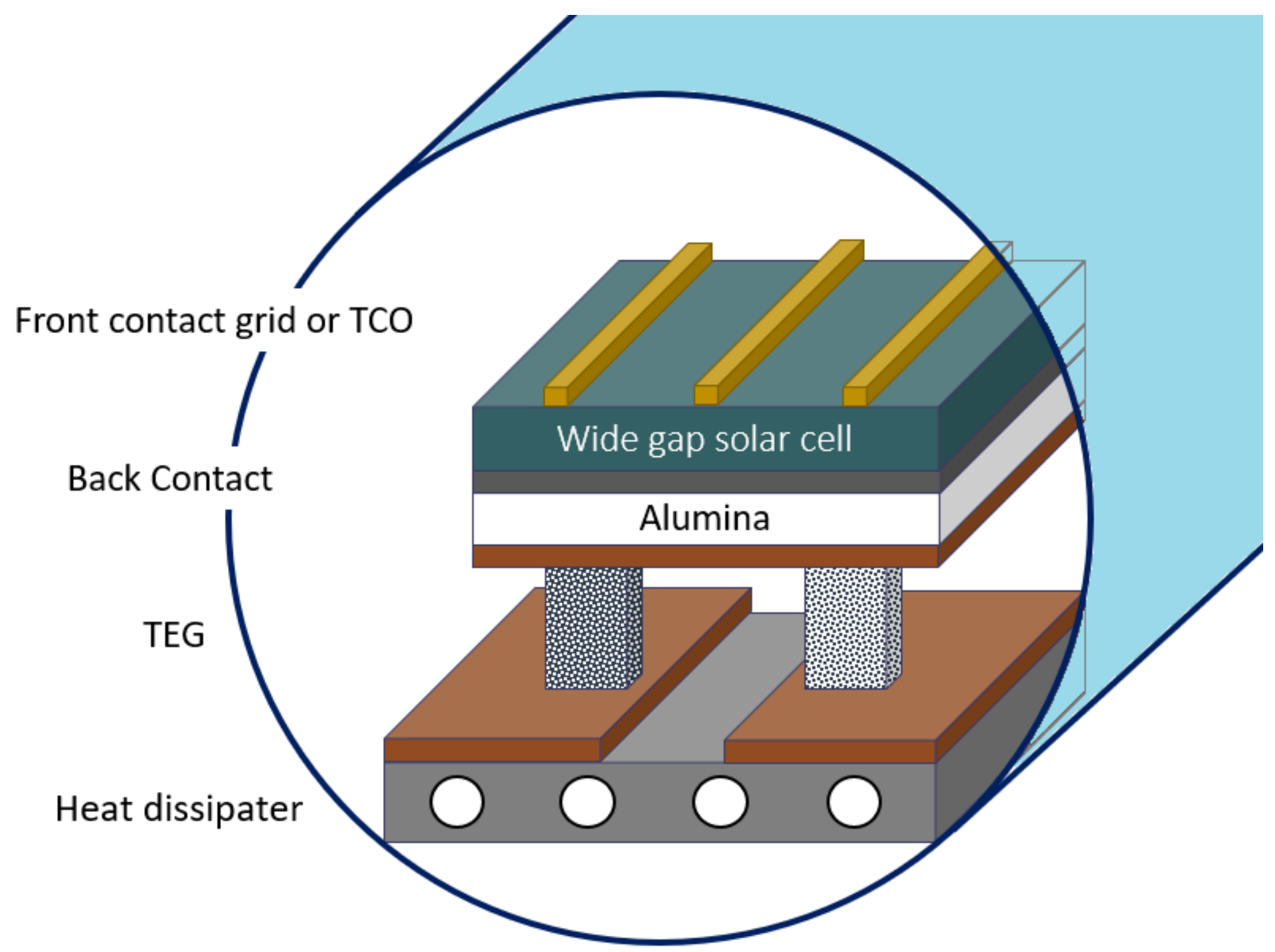

*Figure 1: Scheme of the HTEPV system object of this work*

For this reason, the temperature sensitivity of the PV part is a key parameter, and the kind of solar cells to be hybridized must be chosen carefully. Temperature sensitive materials, such as silicon, loose too much efficiency to make the hybridization convenient. On the other hand, wide-gap solar cells are less temperature sensitive so that their thermoelectric hybridization might be convenient, as shown in our previous work [32]. The physical reason behind this behavior relies on the fact that temperature sensitivity depends on the electron-hole recombination rate. The wider the energy gap, the smaller the recombination rate, and consequently the temperature sensitivity. For a detailed discussion on this matter the reader may refer to [33].

Another consideration deals with the electrical connection between the PV and TEG parts. The case in which the two components are electrically separated is the simpler. For this kind of systems the thermal flux flowing through the device is the main parameter on which designing the HTEPV components. For the case of electrical hybridization, instead, also the internal resistance of the PV and TEG parts must be considered, and a current matching optimization must be considered. Several works have been devoted to electrically hybridized systems with the conclusion that in general it is possible to match the two components without having additional losses [15,34]. In this perspective, the case in which the PV and TEG parts are separated is a simplification of the fully hybridized case and it is the one implemented in this paper.

Summarizing, in this work we developed a thermally coupled, electrically separated, HTEPV device based on wide-gap solar cells. For such kind of systems the hybrid efficiency can be written as

$$\eta_{htepv} = \eta_{pv} + \eta_{teg}\,\eta_{ot} \tag{1}$$

where $\eta_{pv}$ and $\eta_{teg}$ are respectively the temperature dependent PV and TEG efficiencies, and $\eta_{ot}$ the so-called opto-thermal efficiency.

The PV efficiency is normally reported to have a linear decrease as a function of temperature [35]. In this work $\eta_{pv}$ is determined by a detailed characterization of the wide-gap solar cells chosen. The TEG efficiency ($\eta_{teg}$) is instead calculated on the material thermoelectric properties as follows[36]

$$\eta_{teg} = \frac{T_h - T_c}{T_h} \frac{\sqrt{1 + Z_{pn}\, T_m} - 1}{\sqrt{1 + Z_{pn}\, T_m} + \frac{T_c}{T_h}} \qquad (2)$$

with $T_h$ and $T_c$ respectively the hot and cold temperatures of the TEG, $T_m = \frac{T_h + T_c}{2}$, and $Z_{pn}$ the thermoelectric figure of merit defined as

$$Z_{pn} = \frac{\left(\alpha_p - \alpha_n\right)^2}{\left(\sqrt{\kappa_p\, \rho_p} + \sqrt{\kappa_n\, \rho_n}\right)^2} \qquad (3)$$

where $\alpha$, $\rho$, and $\kappa$ are the Seebeck coefficient, the electrical resistivity, and the thermal conductivity of the p and n thermoelectric materials. In this work the TEG was developed starting from commercial bismuth telluride ($Bi_2Te_3$) wafers, and the thermoelectric properties were given by the producer. Once the TEG was built its electrical properties were then characterized and compared with the producer's data.

The opto-thermal component $\eta_{ot}$ of Eq. 1 is instead the efficiency with which the HTEPV system converts the incoming power into heat flowing through the thermoelectric part. Actually, not all the solar power is converted into heat flowing through the TEG, as some of it is reflected by the solar cell top surface, and part of the heat generated within the solar cell is lost towards the environment. In the present work we consider placing the HTEPV device encapsulated within an evacuated environment. Thus, only radiative losses have to be taken into account. In this perspective the opto-thermal efficiency can be written as [37]

$$\eta_{ot} = \left[(1 - \eta_{pv})(1 - R_{pv})\tau_{enc}\right] - \frac{Q_{rad}}{P_{in}} \qquad (4)$$

with $R_{pv}$ and $\tau_{enc}$ respectively the solar cell optical reflectance and the encapsulation optical transmittance, and with $P_{in}$ and $Q_{rad}$ respectively the solar input power and the sum of all the radiative heat losses. In Eq. 4 the product of the terms in the square brackets is the efficiency with which the solar cell converts the input solar power into heat. We assume therefore that all the power not becoming electrical output within the solar cell, excluding optical losses (namely $R_{pv}$ and $\tau_{enc}$) becomes heat. This assumption is justified on the basis of estimations of heat losses available in solar cells [9].

The input power can be written instead as

$$P_{in} = \gamma_{opt}\, G\, A_{abs} \tag{5}$$

where $\gamma_{opt}$, $G$, and $A_{abs}$ are respectively the optical concentration, the input solar power density, and the solar cell active area.

Heat losses ($Q_{rad}$) account for radiative losses between the solar cell top surface and the environment, and between the TEG hot and cold sides. Following the model of our previous study [37], it can be shown that

$$Q_{rad} = 2\, \varepsilon_{eff}\, \sigma\, A_{abs}\left(T_h^4 - T_c^4\right) \tag{6}$$

where $\sigma$ is the Stefan-Boltzmann constant and $\varepsilon_{eff}$ is an effective emittance accounting for both sources of radiative loss. Eq. 6 is a simplification that can be made by taking some assumptions. Actually, the radiative losses from the PV top surface should be written as

$$Q_{rad}^{pv} = \varepsilon_{pv}\, \sigma\, A_{abs}\left(T_{pv}^4 - T_a^4\right) \tag{7}$$

where $\varepsilon_{pv}$ is the top surface solar cell emittance, $T_{pv}$ the PV temperature, and $T_a$ the environment temperature; while the radiative loss across the TEG is

$$Q_{rad}^{teg} = \varepsilon_{teg}\, \sigma\, \left(A_{teg} - A_{te}\right)\left(T_h^4 - T_c^4\right) \tag{8}$$

where $A_{teg}$ is TEG area, $A_{te}$ is the area occupied by thermoelectric legs, and $\varepsilon_{teg}$ is the effective TEG emittance accounting for radiative exchange from the hot to the cold plates, and from the cold to the hot plates of the TEG. It can be shown that [38]

$$\varepsilon_{teg} = \frac{1}{\dfrac{1}{\varepsilon_h} + \dfrac{1}{\varepsilon_c} - 1} \tag{9}$$

with $\varepsilon_h$ and $\varepsilon_c$ respectively the emittances of the hot and cold TEG plates.

Equations 7 and 8 can be joined to give Eq. 6, if

1. the PV and TEG parts have the same areas ($A_{abs}$=$A_{teg}$)
2. the total TEG area is much bigger than the area occupied by thermoelectric legs ($A_{teg}$>>$A_{te}$)
3. the PV have the same temperature of the TEG hot side ($T_{pv} = T_h$)

4. the TEG cold side is at room temperature ($T_c = T_a$)

If all these assumptions are satisfied, it follows that $A_{abs}$ = $A_{teg}$ = $A_{teg} - A_{te}$, and $T_{pv}^4 - T_a^4$ = $T_h^4 - T_c^4$, and Eqs. 7 and 8 can be joined. In the present work all the above assumptions are taken, since they are well satisfied by the experimental setting and the system characteristics. Therefore, in the present case the effective device emittance is simply the average of the TEG and PV emittances $\left(\varepsilon_{eff} = \frac{\varepsilon_{pv}+\varepsilon_{teg}}{2}\right)$ and radiative losses are calculated by Eq. 6.

It also useful to introduce the so-called TEG filling factor as

$$FF = \frac{A_{te}}{A_{teg}} \qquad (10)$$

Note that assumption 2 implies that the TEG filling factor is small. In view of assumption 1, it follows then that $FF = \frac{A_{te}}{A_{abs}}$.

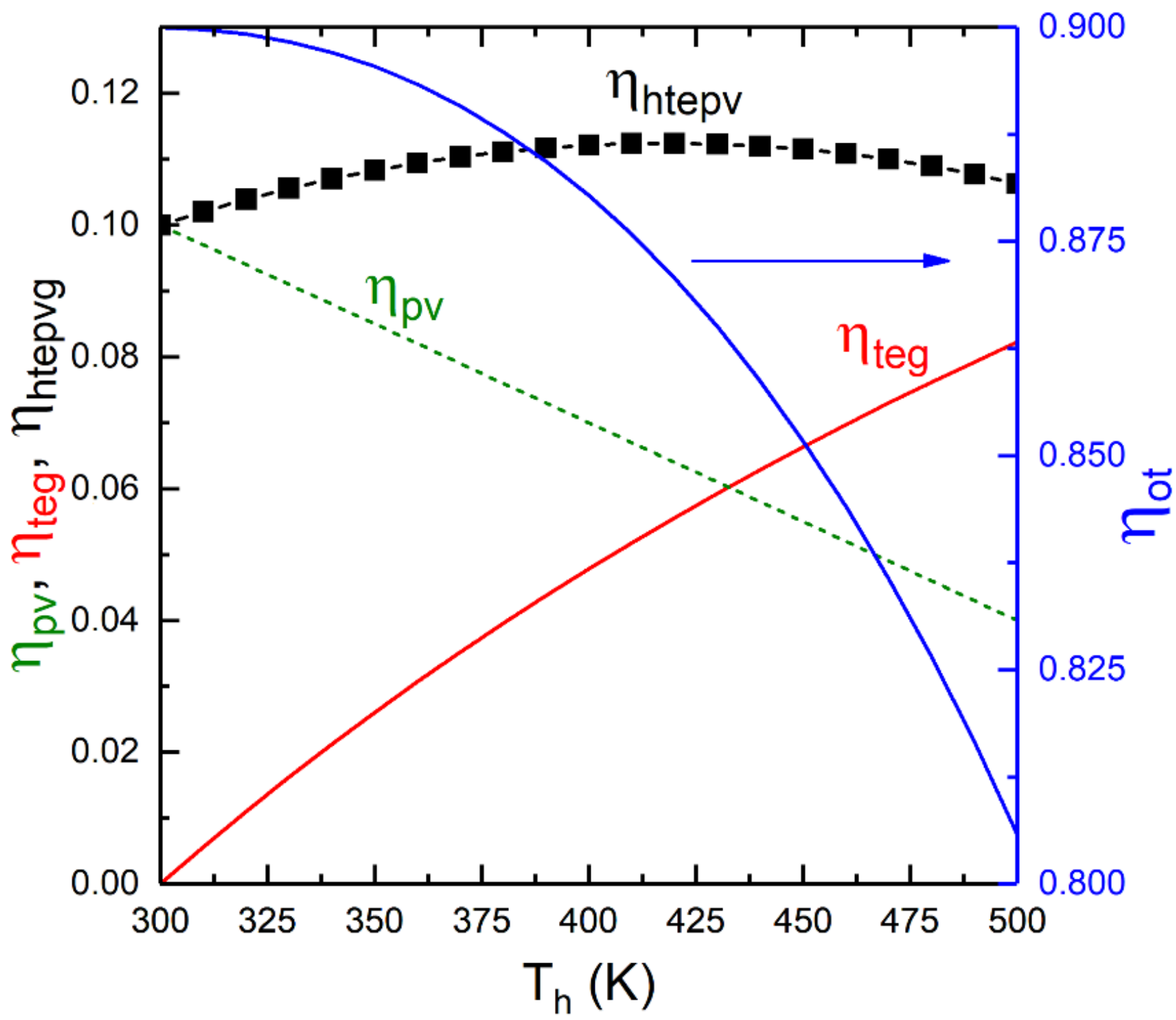

*Figure 2: Generic trends of $\eta_{htepv}$, $\eta_{pv}$, $\eta_{teg}$, and $\eta_{ot}$ as a function of $T_h$*

In Fig. 2 the typical behaviors of $\eta_{htepv}$, $\eta_{pv}$, $\eta_{teg}$, and $\eta_{ot}$ as a function of $T_h$ are reported. It can be seen in the graph that while the PV and opto-thermal efficiencies decrease versus temperature, the TEG efficiency increases. This sets an optimal temperature of operation for the HTEPV system ($T_{opt}$).

This is a key point for the engineering of the device. Actually, once the PV efficiency versus temperature and its optical characteristics ($R_{pv}$, and $\varepsilon_{pv}$) are determined by characterization, and once also the TEG material properties versus temperature are defined, it is possible to predict the behavior of $\eta_{htepv}$ as a function of $T_h$ and determine $T_{opt}$. Based on the optimal working temperature the TEG leg sizes can be computed. In fact, the heat flowing through the TEG can be written as

$$Q_{te} = (\alpha_p - \alpha_n)T_h I + \frac{T_h - T_c}{R_p^{th} + R_n^{th}} - \frac{I^2\left(R_p^{el} + R_n^{el}\right)}{2} \qquad (11)$$

with $I$ the TEG electrical current, $R_p^{th}$ and $R_n^{th}$ the p and n thermoelectric materials thermal resistances, and $R_p^{el}$ and $R_n^{el}$ the p and n thermoelectric materials electrical resistances. The first, second, and third addend of Eq. 11 are respectively the Peltier heat, the conduction heat (Fourier's law), and Joule's heat.

The TEG leg sizes are represented into Eq. 11 by the following relationships:

$$R_p^{th} = \frac{L_p}{\kappa_p\, A_p}, \;\; R_n^{th} = \frac{L_n}{\kappa_n\, A_n} \qquad (12)$$

$$R_p^{el} = \frac{\rho_p L_p}{A_p}, \;\; R_n^{th} = \frac{\rho_n L_n}{A_n} \qquad (13)$$

where $L_p$, $L_n$, and $A_p$, $A_n$ are respectively the p and n leg lengths, and cross-sectional areas. For the sake of simplicity in this work we will consider a TEG generator composed by only one p-n couple of thermoelectric materials, with the same lengths (namely $L_p$ = $L_n$= $L$). It follows that $A_{te} = A_p + A_n$.

Since the opto-thermal efficiency defines the efficiency with which the HTEPV system converts the incoming power into heat flowing through the TEG it follows that $Q_{te}$ con be also be written as

$$Q_{te} = P_{in}\, \eta_{ot} \qquad (14)$$

Therefore, once $T_{opt}$ is defined, it is possible to compute $\eta_{ot}$ from Eq. 4, and then calculate the optimal leg dimensions using Eqs. 11, 12, 13, and 14 in order to keep the system working at optimal temperature. In the calculation it has to be taken into account that when the p and n legs have the same thickness an optimal ratio between the n and p leg areas exists, and it is equal to

$$\gamma_{np}^{opt} = \frac{A_n}{A_p} = \sqrt{\frac{\rho_n \kappa_p}{\rho_p \kappa_n}} \qquad (15)$$

Eq. 15 is valid not only for the HTEPV case, but it is a general argument firstly proposed by Ioffe [39] applicable to TEG generators when operating between two heat sinks.

Finally, it is useful to define the efficiency gain, hereafter named Energetic Convenience Index ($EnCI$), due to thermoelectric hybridization as

$$EnCI = \eta_{htepv} - \eta_{pv}^{0} \tag{16}$$

where $\eta_{pv}^{0}$ is the PV efficiency at room temperature. Thus, $EnCI$ enables a quantitative comparison between HTEPV systems and the best PV scenario, namely cooled solar cells.

In summary, we firstly carried out a detailed characterization of three kind of wide-gap solar cells, in order to determine their efficiency dependence over temperature and their optical characteristics ($R_{pv}$, and $\varepsilon_{pv}$). Then the results of this characterization are inputted along with the thermoelectric material properties in Eq. 1 in order to calculate the HTEPV efficiency behavior versus temperature and to find the optimal working temperature $T_{opt}$. Then using Eq. 14 the heat flowing through the TEG at $T_{opt}$ is calculated and using Eqs. 11, 12, 13, and 15, the thermoelectric leg areas $A_n$ and $A_p$ are found for a given value of the leg thickness $L$. Finally using Eq. 10 the value of FF can be determined.

# 3 Results and discussion

*Solar cell description and characterization*

In this work the solar cells chosen to be hybridized are based on aSi, InGaP, Perovskite. aSi and InGaP solar cells were acquired respectively from the Italian National Energy Agency (ENEA) [40] and from the National Renewable Energy Laboratory (NREL) [41]. Perovskite solar cells (PSCs) with a direct (n-i-p) mesoporous structure, were instead developed at the University of Rome Tor Vergata.

The structure of mesoporous PSCs used in this work is displayed in left. A compact-$TiO_2$, acting as a blocking layer for hole injection into the FTO electrode was deposited *via* spray pyrolysis on a glass/FTO (Fluorinated Tin Oxide). The mesoporous-$TiO_2$ layer, deposited *via* spin coating and sintered at 480 °C, acts instead as scaffold layer for the perovskite growth. Both the compact-$TiO_2$ and the mesoporous $TiO_2$ form the Electron Transport Layer (ETL). A triple cation perovskite ($Cs_{0.05}FA_{0.8}MA_{0.15}Pb(I_{0.85}Br_{0.15})_3$) absorber was deposited *via* spin coating over the mesoporous-$TiO_2$ layer, using Chlorobenzene as antisolvent, with one step deposition technique [42]. Just after its deposition, perovskite was annealed at 100 °C, and a SpiroMeOTAD Hole Transport Material (HTM) was deposited *via* spin coating over it. Then, a Au back electrode was deposited by thermal deposition (i.e. vacuum evaporation at $10^{-6}$ bar).

In this work, the glass substrate was 25 x 25 mm$^2$ and the FTO was etched (1 mm in width, see right) by means of a Nd:$YVO_4$, λ= 1064 nm raster scanning laser, in order to separate the anode and the cathode. The active area for the measurements of electrical parameters (PCE, $J_{SC}$, FF and $V_{OC}$) was 1 cm$^2$.

To protect the PSC from degradation due to oxygen and moisture [43], an encapsulation strategy similar to that reported by Matteocci et. al [44] has been used.

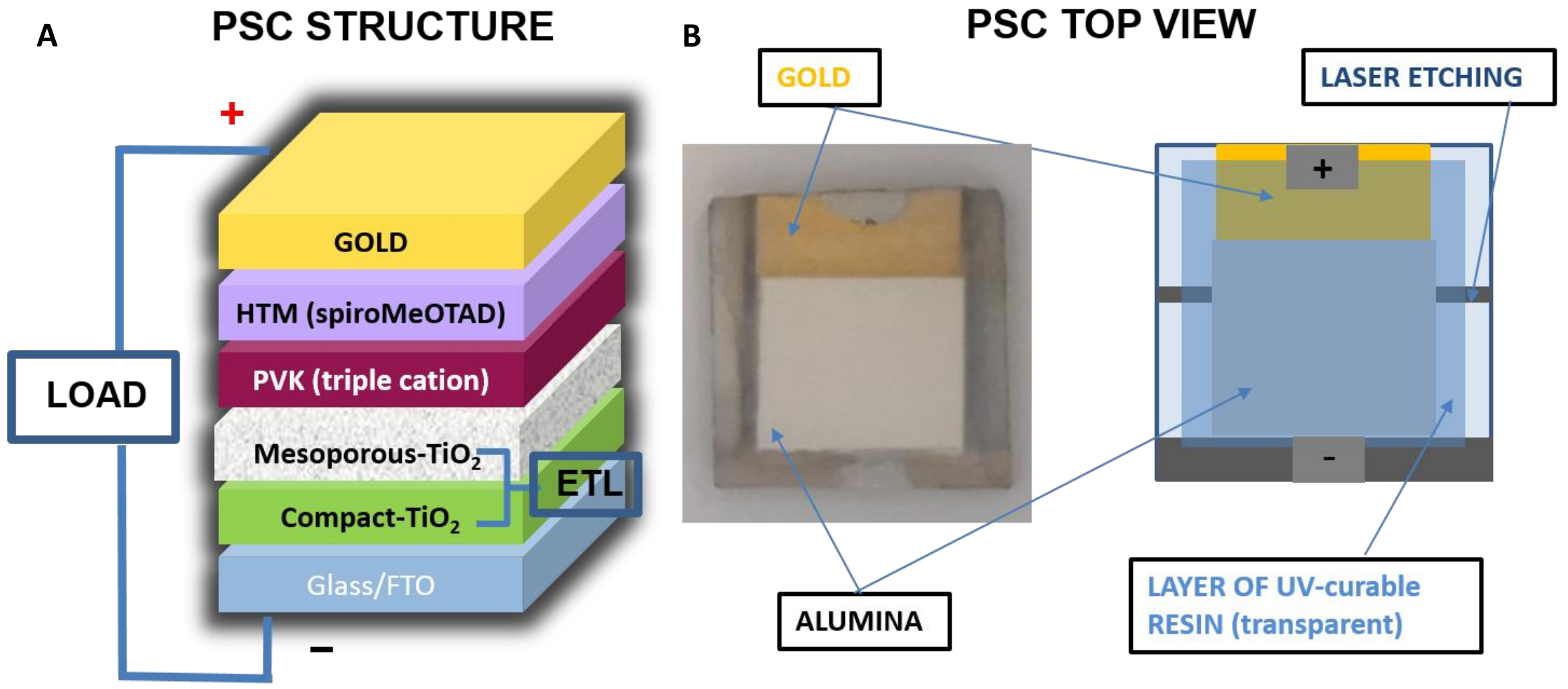

*Figure 3: Structure of the perovskite cells developed in this work*

*(A) Structure of our mesoporous PSC. From bottom to top: conductive glass substrate (2.2 mm) covered with a FTO layer (around 700 nm), compact $TiO_2$ layer (around 50 nm), mesoporous $TiO_2$ layer (around 200 nm), perovskite layer (around 500 nm), spiroMeOTAD layer (300 nm), gold layer (around 100 nm).*

*(B) Top view of the PSC. A layer of UV-curable resin covers the area surrounding the alumina sheet that in turn covers the gold layer in correspondence of the active area of the PSC (located between the laser etching and the silver printed on the anode).*

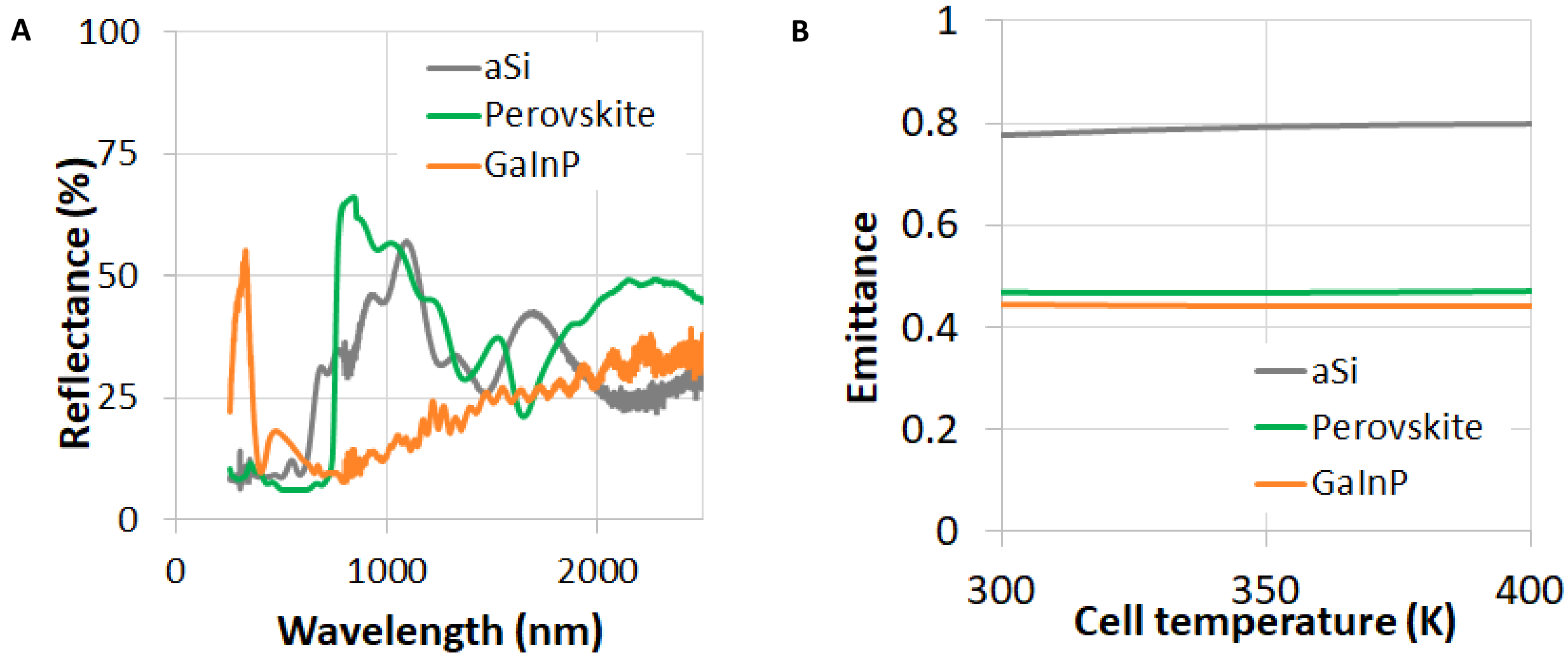

*Figure 4: Optical characterization of the solar cells considered in this work*

*(A) UV-vis reflectance spectra of the three cells considered.*

*(B) Emittances calculated on IR reflectance spectra of the three cells considered.*

As reported in the previous section, the characterization needed to predict the HTEPV performances consists in the measurement of solar cells reflectance, emittances, and temperature sensitivity.

Diffuse reflectance spectra are reported in Fig. 4. Emittance was instead estimated by infrared reflectance as

$$\varepsilon(T) = \frac{\int BB(\lambda,T)\,[1-R(\lambda)-\tau(\lambda)]\,d\lambda}{\int BB(\lambda,T)\,d\lambda} \qquad (17)$$

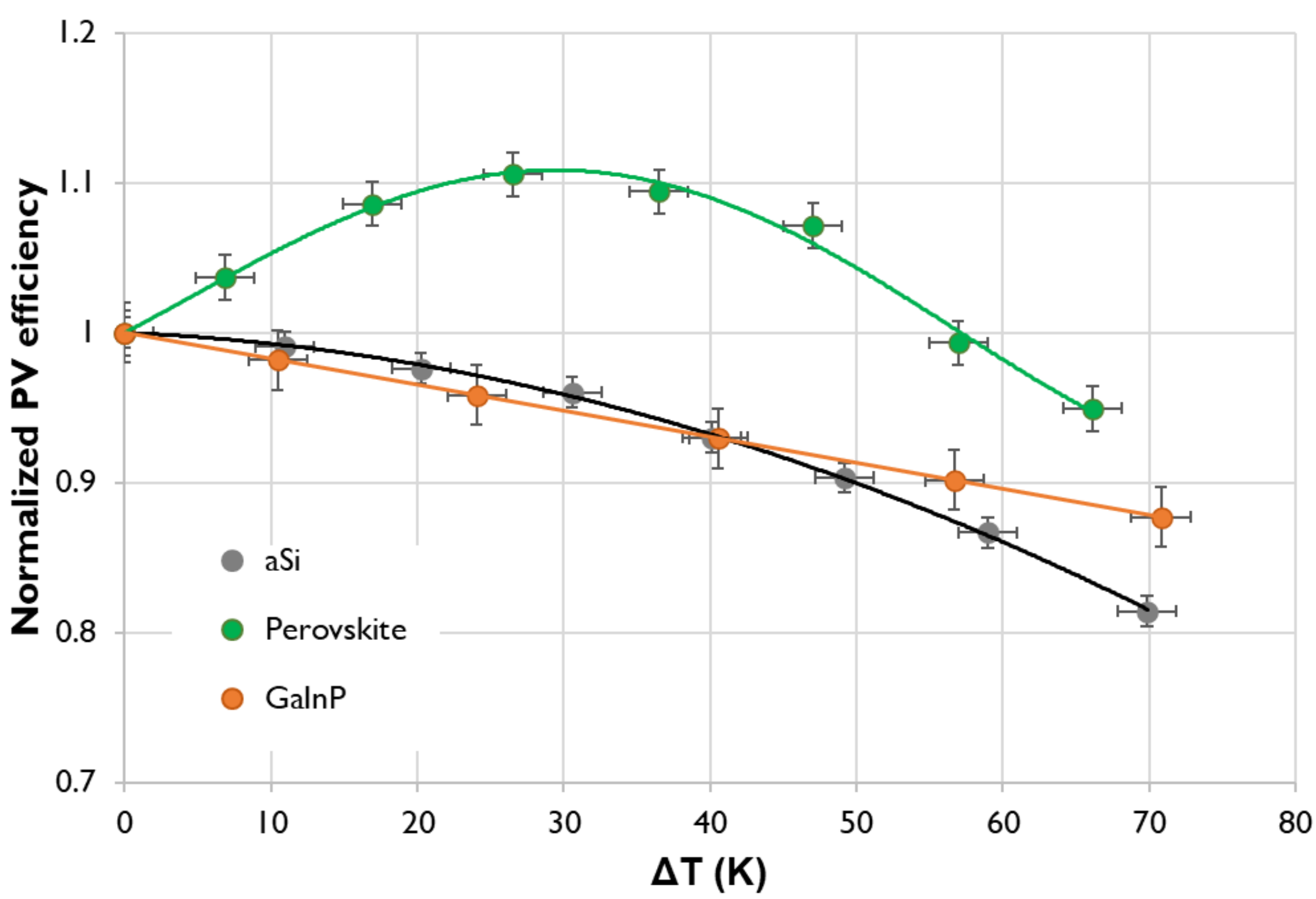

***Figure 5: Measured efficiency versus temperature at 1 Sun, normalized at their value at room temperature.***

with $BB(\lambda, T)$ the calculated spectral power density of a black body at given temperature $T$ and wavelength $\lambda$, while $\tau(\lambda)$ and $R(\lambda)$ the measured transmittance and reflectance at the given wavelength. It has to be noted that for all the solar cells considered in this work $\tau(\lambda) = 0$, since UV-Vis and IR transmittances were found to be null. Furthermore, in Eq. 17, $R$ depends only on $\lambda$ since the reflectance was taken at room temperature. Thus, in this evaluation we assumed a temperature independent reflectance. Fig. 4 shows emittance values for the three solar cells considered.

The results of the temperature sensitivity characterization performed on the solar cells at 1 sun are reported in Fig. 5. As the graph displays, while GaInP shows the expected linear decrease versus temperature, both aSi and perovskites exhibit a nonlinear behavior. Especially for the latter the efficiency shows an interesting increase, reaching a maximum in the temperature range 45 – 55 °C, followed by a strong decrease. This behavior is consistent with what reported in literature [45], and coherent with a tetragonal-to-cubic phase transition reported for MAPI perovskites [46–48]. Also regarding the cases of aSi [49] and GaInP [41], results are in line with literature data. For perovskites, Fig. 5 reports only the case of stabilized efficiencies, which are more reliable than the reverse and forward sweeps [50]. A detailed summary of the characterization on the perovskite and the other solar cells implemented in this work are reported in the Supplementary Information (SI). Finally, the temperature sensitivity of the three solar cells were fitted by polynomials (full lines in Fig. 5).

*Thermoelectric material*

The HTEPV system developed in this work is intended for low optical concentration levels. The reasons for this choice are several. First, low concentration implies optimal operation temperature typically under 400 K. Limited working temperatures are beneficial in terms of materials stability (especially for the solar cell), allow the use of only one thermoelectric material for the TEG part (avoiding the difficulties related to stacked or segmented TEGs), and keep the system in a temperature regime where other hybrid PV/thermal technologies are difficult to be implemented.
Secondly, at low concentration levels the system is more scalable, with limited occupational areas for the optics, with in addition the interesting possibility to avoid daily tracking (e.g. using Compound Parabolic Concentrators - CPCs). Finally, also regarding the dissipation at the TEG cold side, small concentration implies small heat flux to be dissipated and thus simplifying the system.
For temperatures between 300 and 400 K the most efficient and reliable thermoelectric material is bismuth telluride. For this reason, wafers of both p and n type $Bi_2Te_3$, 1 mm thick, coated with titanium were acquired. Producer's data regarding the wafers thermoelectric properties are reported in the SI.

*Predicted efficiency gains*

Fig. 6 summarizes the results of the model based on the PV characterization described in the previous section. The graphs compare the efficiencies of the PV (dashed lines) and the HTEPV (full lines) as a function of temperature and optical concentration for the three PV materials considered in this paper. As anticipated, for all three cases, thermoelectric hybridization leads to efficiencies higher than the efficiency of the sole PV case at room temperature. These gains show partially similar trends among materials. In all cases, the higher the optical concentration, the higher the gain. This behavior is due to a double beneficial effect of concentration, which on one hand increases the PV efficiency and mitigates the PV temperature sensitivity, and on the other hand reduces the effect of radiative thermal losses. The first effect is due to a higher carrier generation/recombination ratio, which increases the PV efficiency and decreases its temperature sensitivity. The second effect is instead due to a higher opto-thermal efficiency. In fact, the increase of optical concentration increases the value of $P_{in}$ in Eq. 4, leaving instead $Q_{rad}$ unaltered.

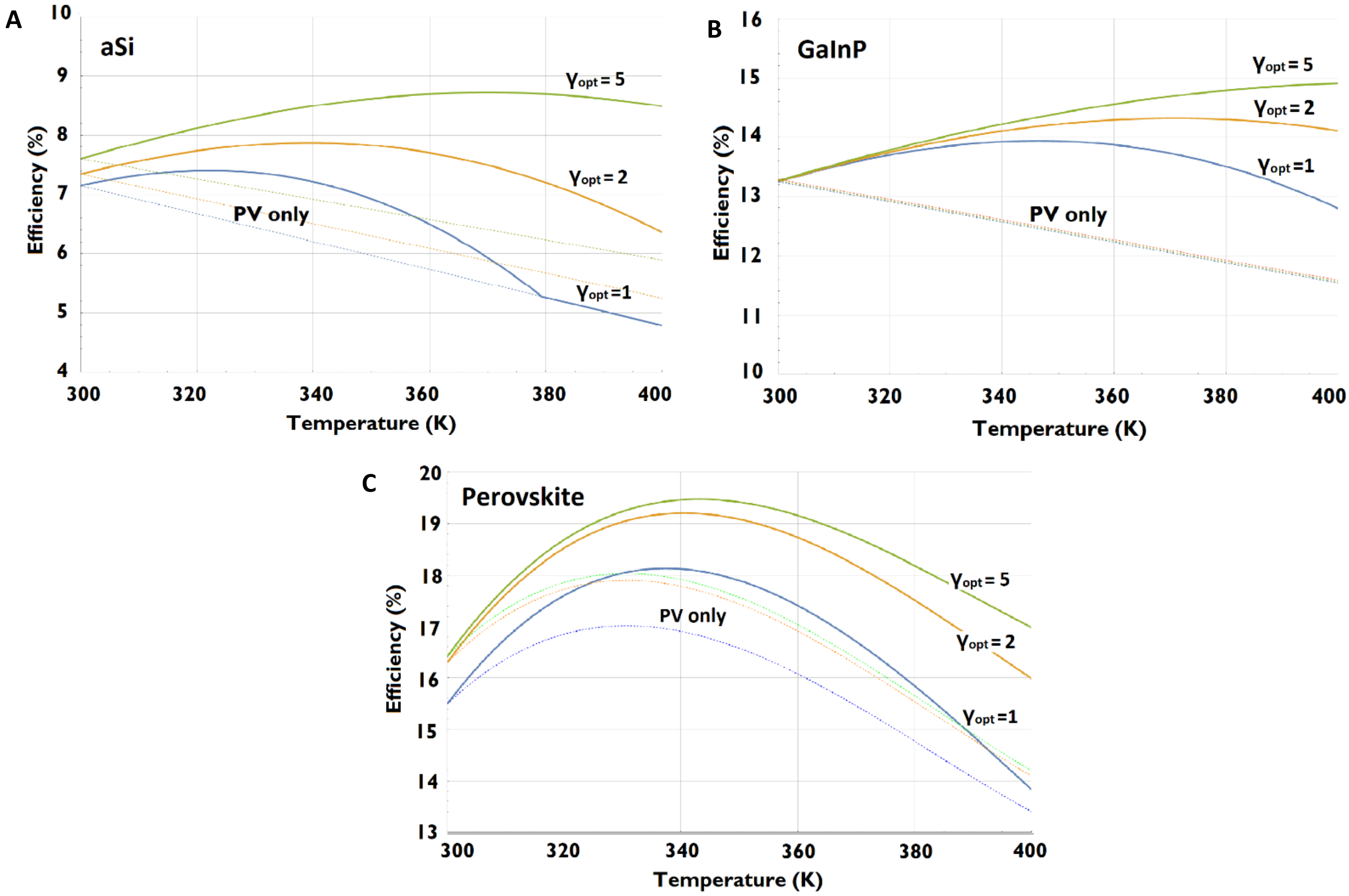

***Figure 6: Predicted $\eta_{pv}$ and $\eta_{htepv}$ as a function of temperature.***

*(A) Amorphous silicon*
*(B) Gallium Indium Phosphide*
*(C) Perovskite*

For the cases of GaInP and aSi the optimal temperature increases with the concentration. At concentration 5× the optical temperature falls in the range 380 – 400 K. For perovskite, instead, the optimal temperature is stable around 340 – 350 K and is independent of the optical concentration. The main reason for this dissimilarity is given by the strong decrease of the perovskite PV efficiency for temperature higher than 340 K.

Fig. 6 summarizes all efficiency gains (*EnCI*, as defined by Eq. 16). In addition, for the case of perovskite, a further gain, defined as the difference between the maximum HTEPV efficiency and the maximum PV efficiency is reported since in this case the PV efficiency at room temperature is not the PV best case scenario.

***Table 1. Predicted EnCI values***

| | $\gamma_{opt} = 1$ | $\gamma_{opt} = 2$ | $\gamma_{opt} = 5$ |
|---|---|---|---|
| aSi | 0.26 %<br>(322.23 K) | 0.53 %<br>(339.42 K) | 1.11 %<br>(370.02 K) |
| GaInP | 0.68 %<br>(346.47 K) | 1.07 %<br>(370.86 K) | 1.64 %<br>(408.29 K) |
| Perovskite | 2.64 %<br>(337.43 K) | 2.90 %<br>(340.59 K) | 3.05 %<br>(343.13 K) |
| Perovskite<br>(gain with respect to the best PV efficiency) | 1.12 % | 1.30 % | 1.44 % |

*Values of EnCI for the three materials considered at 1×, 2×, and 5× optical concentrations. Temperatures between rounds brackets report the temperature at which the maximum EnCI occurs*

Table 1 clearly shows how perovskite gives the higher values of *EnCI* at all optical concentrations. This evidence, joined with the fact that perovskite has the higher final efficiency (as shown in Fig. 6) led to the choice of this material for the practical demonstration of solar cell hybridization.

*3.3 Optimized TEG development*

As mentioned above, in this work the hybrid system was implemented and fully validated for the case of perovskite. There are multiple reasons for this choice. Firstly, perovskite leads to the highest values of *EnCI* and of the final efficiency. Secondly, the model predicts an optimal temperature around 340 K. This temperature is well within the range of temperatures normally experienced by solar panels and does not imply the need of difficult thermal management strategies. Thus, in this case the HTEPV device is directly comparable and compatible with actual solar cells. Using the model described in Sect. 2 the TEG filling factor and consequently the p and n leg areas were calculated. Table 2 reports the values of the filling factor for optical concentrations of 1, 2 and 5. As shown, filling factors are very small, justifying assumption 2 made in Sect. 2.

**Table 2. Optimal filling factors**

| | $\gamma_{opt} = 1$ | $\gamma_{opt} = 2$ | $\gamma_{opt} = 5$ |
|---|---|---|---|
| Filling factor<br>$(A_{te}/A_{abs})$ | 0.26 %<br>(322.23 K) | 0.53 %<br>(339.42 K) | 1.11 %<br>(370.02 K) |

*Optimized TEG filling factor at different values of optical concentration.*

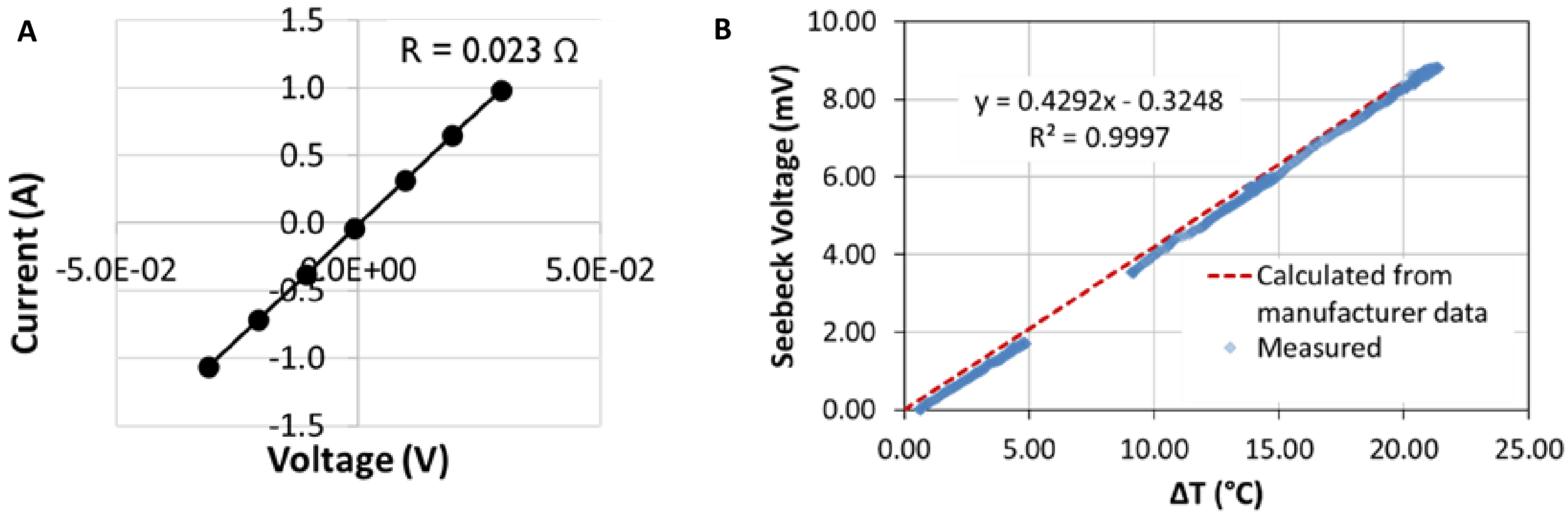

*Figure 7: TEG electrical characterization*

*(A) Dark IV curve for the TEG developed to work at 1 Sun.*

*(B) Measured Seebeck voltage versus the difference of temperature at the cold and hot plates of the TEG developed to work at 1 Sun, compared with the Seebeck voltage calculated on manufacturer's data.*

Based on the model results, $Bi_2Te_3$ wafers were then cut and soldered on copper plates. Finally, the TEGs so obtained were characterized, measuring their internal resistance and their thermoelectric voltage. Fig. 7 left reports the I-V curve for the determination of the TEG internal resistance, while Fig. 7 right shows its thermoelectric response. For the sake of clarity Fig. 7 reports only the case of the TEG optimized to work at no optical concentration. The characterization performed showed that while Seebeck voltages were consistent with manufacturer's specifications, TEG internal resistance is slightly higher than the calculated value (0.018 Ω for the case reported in Fig. 7 left). This suboptimal result is presumably due to the existence of a contact resistance at the leg/plate interface, which in turn leads to a final HTEPV output power smaller than its calculated value.

*HTEPV characterization*

Once the TEG was characterized, the perovskite solar cell was thermally coupled to the TEG hot plate by a layer of thermal grease. The grease was used to allow good thermal contact between the two components

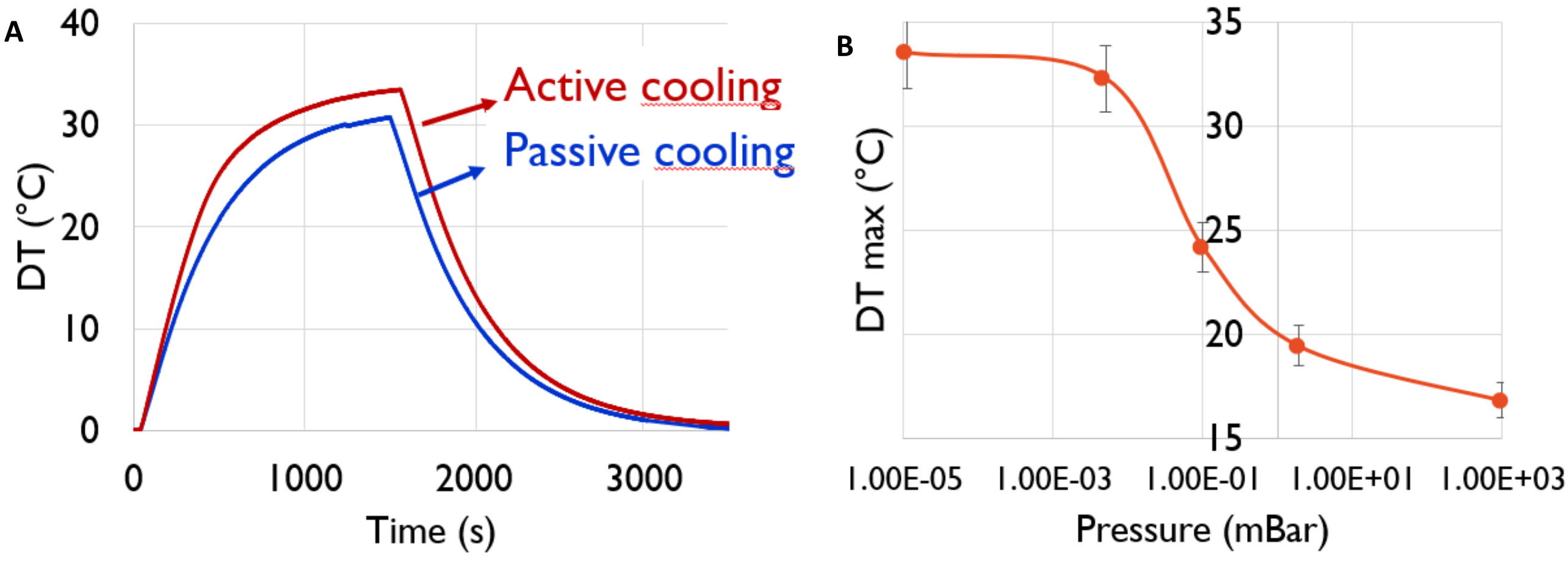

*Figure 8: Effect of active cooling and ambient pressure*

*(A) Effect of active cooling and chamber pressure on the maximum ΔT reached for the HTEPV system based on perovskite at 1 Sun.*

*(B) Effect of the chamber pressure.*

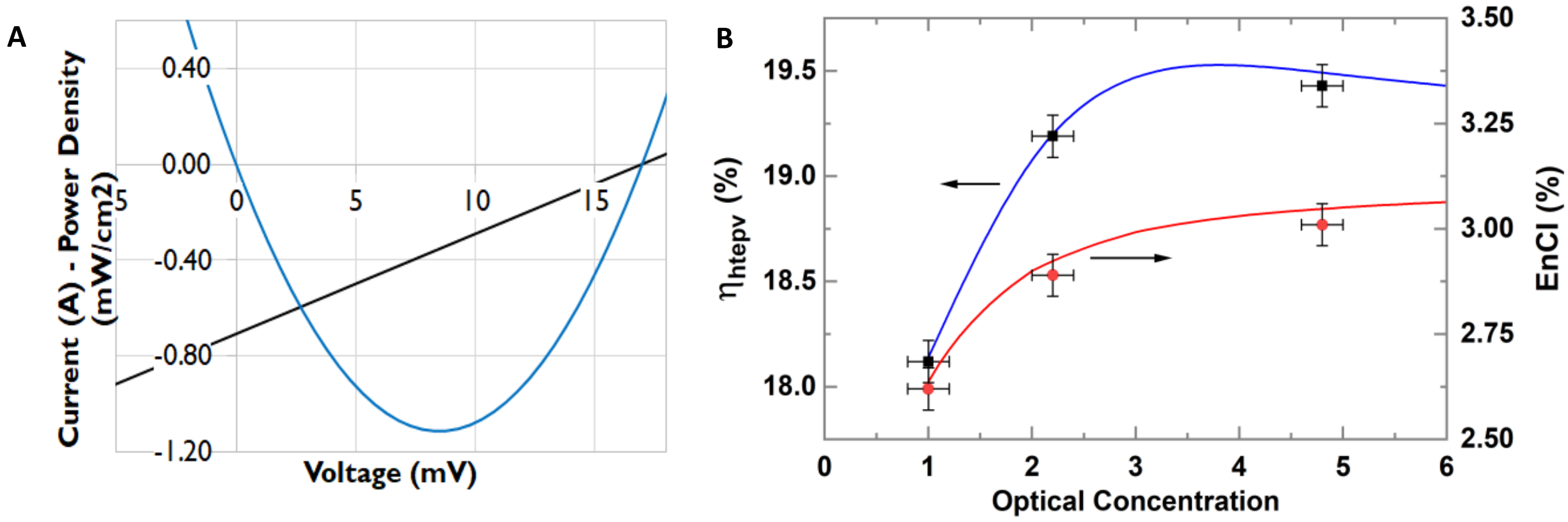

***Figure 9: Electrical characterization of the hybrid device under illumination***

*(A) TEG IV curve (black line) and the output power (blue line) under illumination at maximum ΔT for the case of HTEPV system based on perovskite at 1 Sun.*

*(B) Measured $\eta_{htepv}$ at three different optical concentrations (black points) compared with model prediction (blue line). Measured EnCI at three different optical concentrations (red points) compared with model prediction (red line).*

but also to allow the insertion of a thermocouple to measure the temperature at TEG/solar cell interface. The HTEPV device was then placed within a vacuum chamber equipped with a window through which shining light on the hybrid prototype. At the chamber basement the temperature could be controlled by a cooling plate connected to a chiller. In case of active cooling the temperature of the base was always kept at 25 °C. Fig. 8 left reports the influence of the active cooling on the build-up of a temperature difference at the TEG sides, in the case of no optical concentration. As expected, active cooling facilitates a more rapid increase of ΔT over time and allows to reach a higher final temperature (333.53 K). This temperature is anyway slightly lower than the optimal temperature predicted by the model (334.64 K). The reason behind this discrepancy is probably due to the non-optimal thermal contact between the solar cell and the TEG hot plate. Actually, the model does not account for the thermal resistances of the alumina plate at the solar cell bottom and the thermal grease. Luckily, the predicted HTEPV efficiency (showed in Fig. 6) presents a region around the optimal temperature where the efficiency does not change significantly. This is a further strong point of HTEPV device based on perovskites.
Fig. 8 right reports instead the dependency of maximum ΔT versus the chamber pressure. It shows that the maximum ΔT is optimal for a range of pressure between $10^{-5}$ and $10^{-3}$ mBar but drops significantly for higher pressures. At ambient pressure the maximum ΔT is nearly half of that at $10^{-5}$ mBar. Consequently, in this case the TEG contribution is a quarter of that at best case.

Finally, the output power of the TEG coupled with the perovskite solar cell under illumination at max ΔT was measured by a J-V characteristic. In Fig. 9 left the JV curve (black line) along with the power output, hereafter $P_{teg}$ (blue line) are reported. As expected the J-V characteristic is linear with an open circuit voltage ($V_{oc}$) equal to the product between the TEG Seebeck coefficient and ΔT. Regarding the output power, the maximum TEG power, hereafter $P_{teg}^{max}$ (which is the minimum of the blue line in Fig. 9 left, since by convention the output power of a power generator is negative) occurs at a $V_{oc}/2$ corresponding to a load resistance equal to the TEG internal resistance.

From the ratio $P_{teg}^{max}/P_{in}$ it is then possible to calculate the TEG efficiency and consequently the value of the hybrid efficiency ($\eta_{htepv}$) and of *EnCI.* Fig. 9 right reports these results for three different value of optical concentrations compared with the model predictions, showing that the measured values of $\eta_{htepv}$ and *EnCI* follow quite accurately the predicted trend, although computations slightly overestimated them. This difference probably originates from a small mismatch of the TEG internal resistance, due to non- optimal electrical contacts between the legs and the copper plates.
Nevertheless, the comparison between the experimental characterization and the calculated data shows that our model is accurate.

## 4 Conclusions

In this work we demonstrated that the thermoelectric hybridization of wide-gap solar cells can lead to final efficiencies higher than the sole PV case. In particular the cases of aSi, GaInP and triple-cation perovskite were taken into account, and firstly analyzed in the framework of a computational model able to predict the final hybrid efficiency. The input data for the model were obtained by means of a detailed characterization of the given solar cells at different optical concentrations, predicting efficiency gains always positive, ranging between 0.2 and 3.1%. The case of perovskites were found of particular interest since showed efficiency gains higher than 2% at all the optical concentrations considered (1, 2, and 5) with maximum efficiency gains happening at moderate temperatures (around 340 K). For these reasons the case of perovskite was chosen for the development of a hybrid device prototype. To this purpose bismuth telluride wafers were acquired and cut in pieces accordingly to the sizes calculated with the model, able to make the system to work at optimal temperature. The thermoelectric legs were then soldered and the HTEPV system developed. The characterization of the system under solar illumination showed gains at optimal temperature very similar to those predicted, validating the model described and its results. The characterization also showed the influence of the pressure in which the system works, and the effect of active cooling at the TEG cold side. A final maximum efficiency gain of 3.05%, at temperatures normally experienced by solar cells, was thus experimentally demonstrated for the case of perovskite, showing the actual potential of hybrid thermoelectric – photovoltaic systems.

**Acknowledgements**

This project has received funding from the European Union’s Horizon 2020 research and innovation program under the Marie Skłodowska-Curie Grant No. 745304.

## EXPERIMENTAL PROCEDURES

*Materials and chemicals*

For the PSC preparation the following compound were purchased: conductive substrates, with a sheet resistance of 7 Ω sq-1 from NSG-Pilkington; acetone, 2-propanol, diisopropoxide bis(acetylacetonate), titanium(IV) diisopropoxide bis(acetylacetonate), Chlorobenzene, tert-butylpyridine (TBP - 26.77 µl/ml), lithium bis(trifluoromethanesulfonyl)imide, (Li-TFSI - 16.6 µl/ml) and cobalt(III) complex (FK209 - 7.2 µl/ml) from Sigma Aldrich; acetylacetone from Merck; Kapton tape and bynel60 from DuPont; $TiO_2$ paste from GreatCell Solar, spiro-OMeTAD (2,20,7,70-tetrakis-(N,N-dip-methoxyphenylamine)9,9’-spirobifluorene) from Borun New Material, UV-curable resin from Threebond.

For the TEG development silver epoxy, copper sheets, and Kapton tape were purchased by RS components, while Bismuth telluride wafers (n and p type) covered by an electro-deposited titanium contact/protective layer (40 nm) were purchased from Wuhan Dongxin Mill Imp&Exp Trade Co., Ltd. Silicone-free thermal grease was acquired from Fischer elektronik GmbH & Co. KG.

*PSC fabrication*

Conductive substrates were cut in 25 x 25 mm$^2$ size and subsequently cleaned brushing their surface without scratching the FTO (Fluorine Tin Oxide) layer, using “Hellmanex” cleaning solution diluted with water in a 2:98 vol/vol ratio. At the end, substrates were rinsed in 2-Propanol and quickly dried with a strong air flow. Just after the cleaning, an UV-$O_3$ treatment was performed by Pro Series Digital UV Ozone System from "Novascan" in order to remove the organic contaminations [51]. The etching of substrates was performed by "BrightSolutions, Luce 40" laser in order to separate PE and CE (see Fig. 3).

Perovskite solar cells (PSCs) were fabricated mainly by following the steps reported in [42], except for the compact $TiO_2$ layer. Just after the UV-$O_3$ treatment, the FTO/glass substrates were transferred to a hot plate and warmed up to 460 °C (within 35 minutes). The substrates were then left for 10 min at 460 °C before depositing the compact $TiO_2$ layer by automatized spray pyrolysis. The sprayed solution was composed by 0.5 mL of acetylacetone and 0.75 mL of titanium(IV) diisopropoxide bis(acetylacetonate), in 11.25 mL of ethanol (EtOH), which is enough to cover an FTO surface of about 12.5 × 7.5 $cm^2$. The hot plate was inserted into an ASC (Automatic Spray Coating) machine (AUREL s.p.a,, Avigliana, Italy). The nozzle of the ASC machine was about 10 cm far from the FTO surface and it was moved according to a serpentine path, for 5 cycles in order to reach around 50 nm in thickness. Then, the substrates were left for 15 minutes at 460 °C before slowly cooling them down to room temperature. Over the compact $TiO_2$ layer, mesoporous $TiO_2$ layer was deposited (out of the glove box) by spin coating technique. In order to fabricate the mesoporous $TiO_2$ layer, we use 30 N-RD $TiO_2$ paste diluted in EtOH (1:5 in weight). The resulting dispersion is left under stirring at least overnight before use it and 120 µl of it are dropped onto each sample, using a speed of 3000 rpm (1500 rpm/s acceleration) for 20 s. Just after its deposition, the mesoporous $TiO_2$ film was dried at 120°C for 15 minutes in order to completely evaporate the solvent (EtOH). Then samples were sintered at 480°C following the multi temperature steps described in[42]. The sintering process allows to decompose the organic binders in the TiO2 paste and to promote electromechanical bonding between the TiO2 NPs [52]. In order to improve the wettability [53] of the mesoporous $TiO_2$ layer, we put the sample under UV lamp 45 min. This treatment is performed by a DYMAX 5000-PC UV Light-Curing Flood Lamp.

The so called "triple cation" perovskite $Cs_{0.05}FA_{0.8}MA_{0.15}Pb(I_{0.85}Br_{0.15})_3$ [54] was solved in a mixture of DMF (N-Ndimethylformamide)-DMSO (dimethyl sulfoxide) 4:1 in weight. The solution, prepared inside Glove Box, was left stirring overnight at room temperature. Then, 100 µl of solution in the middle of substrate was dropped. The deposition was split in two spinning steps. The first step (10s at 1000 rpm with an acceleration of 200 rpm/s) has the aim to ensure full surface coverage, while the second step (30 s at 5000 rpm with an acceleration of 2500 rpm/s) has the task to prepare the solution for the dropping of the antisolvent. Indeed, 150 µl of Chlorobenzene (CB) were dropped in the middle of the rotating substrate around 10 second before the end of spinning. Just after the deposition, the perovskite layer is immediately placed onto a hot plate at 100°C for the annealing process (for 1 hour). A solution of spiro-OMeTAD (2,20,7,70-tetrakis-(N,N-dip-methoxyphenylamine)9,9'-spirobifluorene, 73.5258 mg/ml was prepared in Chlorobenzene and left stirring overnight. The solution was then doped with tert-butylpyridine (TBP), lithium bis(trifluoromethanesulfonyl)imide (Li-TFSI) and cobalt(III) complex (FK209) and spin coated at 2000 rpm for 20 seconds to form the Hole Transporting Material (HTM). The final step was the deposition of 90 nm-thick Au back electrode. This gold layer is fabricated by means of thermal deposition (i.e. vacuum evaporation at $10^{-6}$ bar).

The encapsulation process is based on the lamination of the alumina sheet onto the adhesive Kapton following Ref. [44] Then, this stack was applied onto the gold layer of PSC (the adhesive side of the tape is in contact with the gold). At the edges of the alumina sheet an UV-curable resin is deposited, in order to further protect the edges of the solar cell active layers form oxygen and moisture.

*TEG fabrication*

Legs with optimized areas were cut from bismuth telluride wafers by an circular saw equipped with a micrometric positioning system, and then adjusted with a polishing machine from Struers Inc. TEG electrodes (two for the cold and one for the hot side) were cut from a 0.45 mm thick polished copper plates. The cold electrodes were then fixed to an alumina support by Kapton tape, and a drop of silver epoxy were placed on them using a toothpick. The thermoelectric legs were positioned on the silver drop and the silver paste cured at 80 °C for two hours in air. Two silver paste drops were then placed on the legs tops, and the hot electrode placed on top of it, to form the typical π TEG shape. Finally the silver paste was cured with the same procedure used for the cold side.

*Measurements and Characterizations*

Solar cell UV-vis reflectance was characterized between 250 and 2500 nm with a Jasco 570 spectrophotometer equipped with an integrating sphere. Infrared reflectance was instead taken between 400 and 4000 $cm^{-1}$, measured with a Thermo Fisher FTIR 6700 equipped with an integrating sphere.

Regarding current-voltage (J-V) characteristics of solar cells and of the hybrid device, the characterization was performed in a vacuum chamber equipped with a window and placed under an ABA, ABET Technologies solar simulator. For the solar cells, they were attached with thermally conductive paste on a controllable heather, and their temperature monitored with K type thermocouples and a Keithley 2700 multimeter. J-V curves were recorded between 25 and 80 °C with steps of ∼10 °C by a Keithley 2440 source meter controlled with a LabView program. Only for the case of perovskites, forward, reverse, and stabilized I-V curves have been recorded, since considerable differences were found among the three cases (details in SI). The characterization was performed for all three cells between 1 and 5 Suns to determine the effect of optical concentration on the temperature sensitivity. The solar simulator incoming power was always measured and adjusted with a certified refence silicon solar cell, and stainless-steel mask with known areas were implemented to accurately evaluate incoming power density. For the final hybrid device, the TEG cold side was attached with thermal grease to the vacuum chamber bottom and its temperature controlled with a K type thermocouple. The chamber bottom temperature was controlled with a dissipation liquid circuit fed with chiller with adjustable temperature. The solar cell was placed in thermal contact to the TEG top electrode by means of a layer of thermal grease, and a K thermocouple placed between the hot electrode and the solar cell bottom. Also in this case, J-V curves were recorded by a Keithley 2440 source meter controlled with a LabView program.

## Supplementary Information

# Practical development of efficient thermoelectric – photovoltaic hybrid systems based on wide-gap solar cells

Bruno Lorenzi,*[1] Paolo Mariani[2], Andrea Reale[2], Aldo Di Carlo[2], Gang Chen[3], Dario Narducci[1]

[1]Dept. Materials Science, Univ. of Milano Bicocca, Milano, 20125, Italy
[2]CHOSE, Center for Hybrid and Organic Solar Energy, University of Rome Tor Vergata, Rome, 00133 Italy
[3]Dept. Mechanical Engineering, Massachusetts Institute of Technology, Cambridge, MA, 02139, USA

## S1 Solar cells IV characterization

*S1.1 Amorphous silicon - aSi*

In Fig. S1 we report the IV curves at different temperatures and optical concentrations of the aSi solar cell characterized in this work. The curves were performed by a forward scan with a scan rate of 666.67 mV/s.

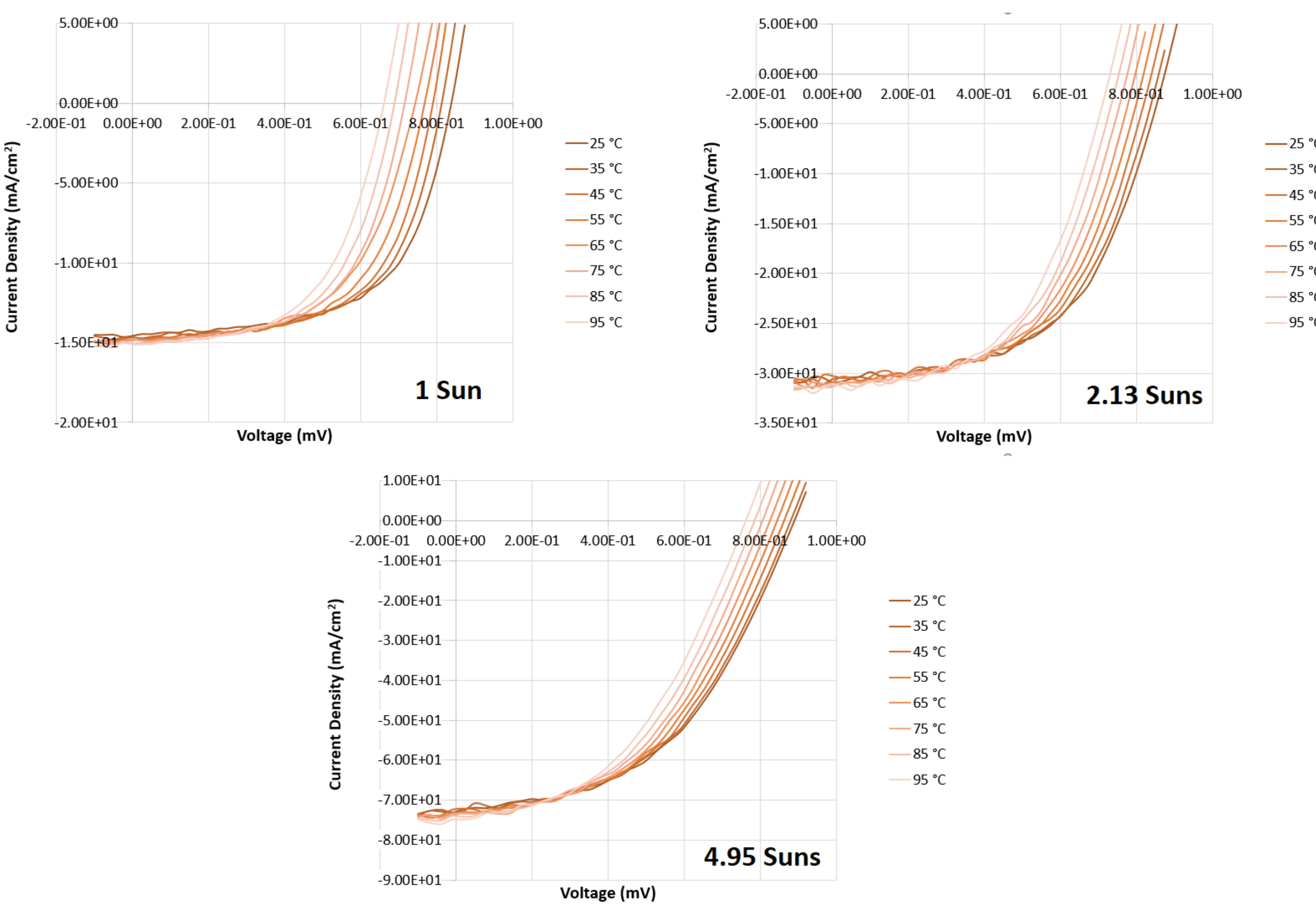

*Figure S1: IV curves at different temperatures and optical concentrations of the aSi solar cell characterized in this work.*

*S1.2 Gallium Indium Phosphide - GaInP*

In Fig. S2 we report the IV curves at different temperatures and optical concentrations of the GaInP solar cell characterized in this work. The curves were performed by a forward scan with a scan rate of 666.67 mV/s.

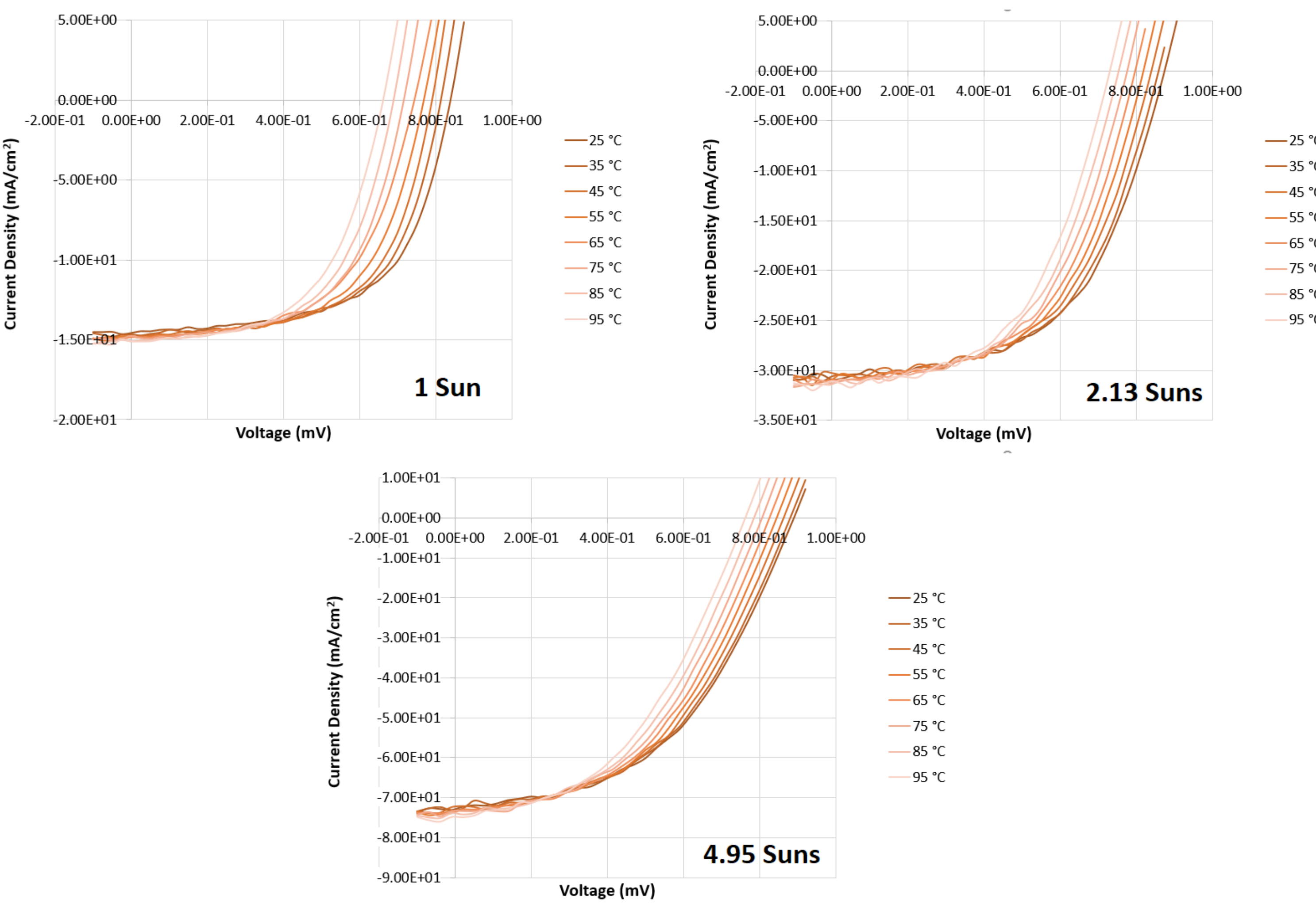

*Figure S2: IV curves at different temperatures and optical concentrations of the GaInP solar cell characterized in this work.*

*S1.3 Perovskite*

It is well known that the electrical behavior of PSCs is very sensitive to the way how they are measured. Their IV characteristics have been shown to depend on the IV sweep settings, such as the direction of the sweep, the scan rate, the presence of a starting bias, and the light soaking time. Consequently the final efficiency and the other device parameters are very sensitive to these settings [1].

In this work we characterized PSCs efficiencies implementing forward, and reverse scans, along with stabilized measures of the output power. Following the approach of Wang et al. [2], forward and reverse sweeps were executed at scan rate of 380 mV $s^{-1}$ with two-wire connection under source mode, but with the difference of a light soaking time of around 15 s. Stabilized efficiencies were instead recorded, by holding the cell at a fixed voltage near the maximum power point for 60 s.

Since for the purpose of this work we were interested on the temperature sensitivity of PSCs, we firstly studied the effect of the kind of measure on it. We found that for all the cases the efficiency increases, with a maximum around 45 – 55 °C, followed by a strong decrease (as shown in Fig. S3).

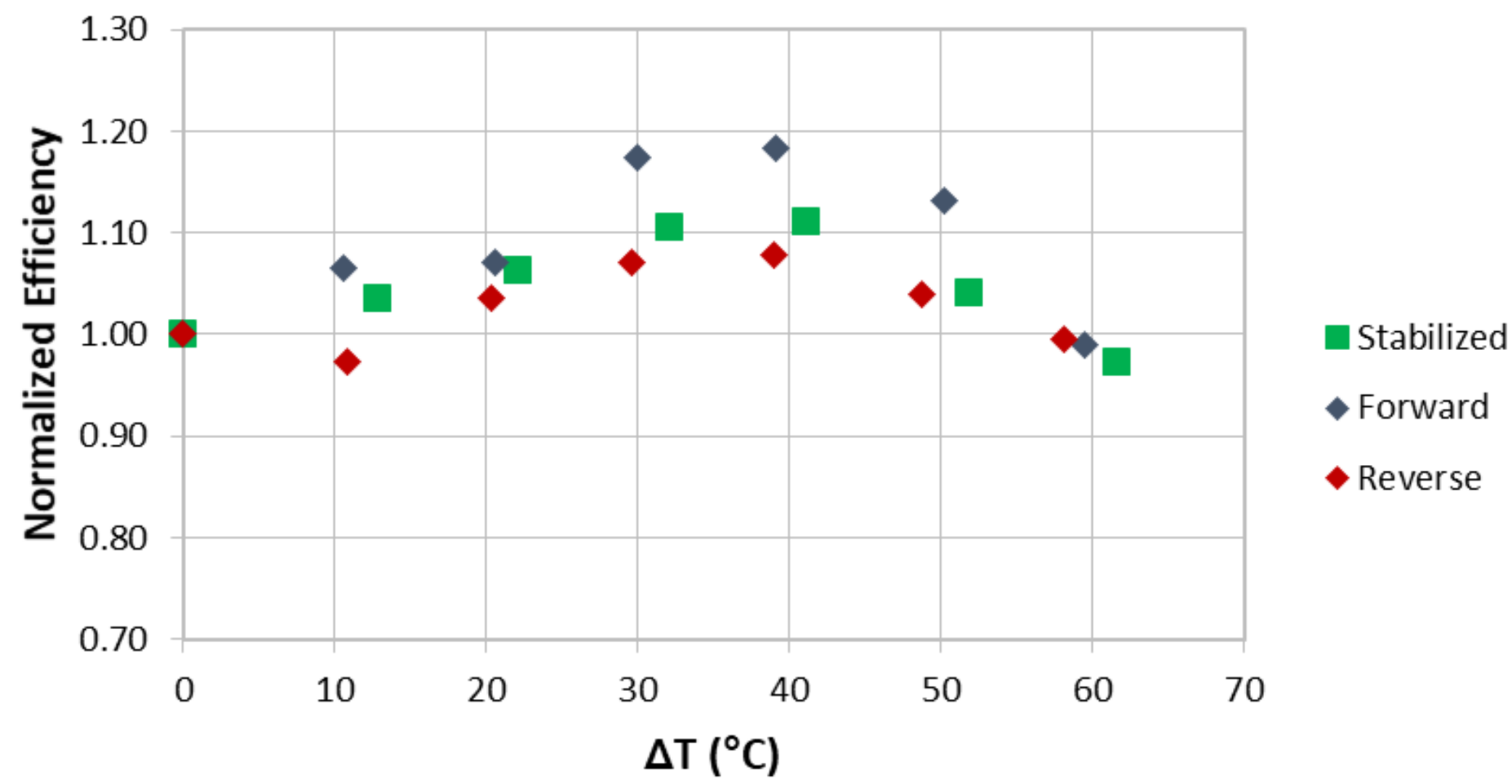

*Figure S3: Normalized efficiency versus temperature in the case of stabilized power output measures, and IV sweeps in forward and reverse directions.*

Typically the forward scan case showed a stronger increase compared to the reverse one, with the stabilized case in between. From this evidence, and from the fact that in general stabilized measures are the most reliable approach to characterize PSCs, we decided to focus our characterization on stabilized measures.
In Fig. S4 we report an example of this kind of characterization at different temperatures.
While in Fig. S5 we report the normalized efficiency versus temperature, at different optical concentration, implemented in the model reported in the main manuscript.

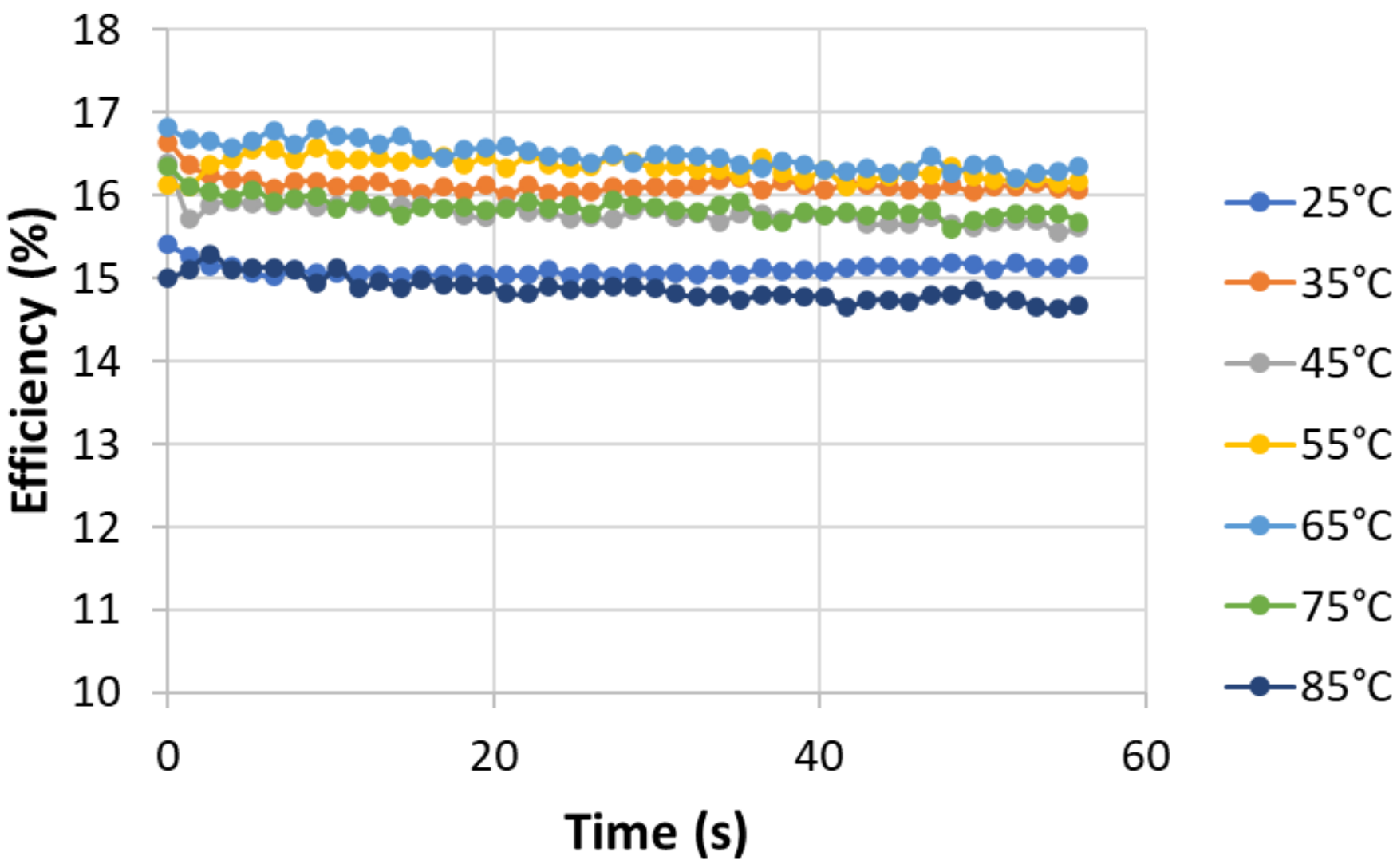

*Figure S4: Stabilized measured of PSCs at different temperatures.*

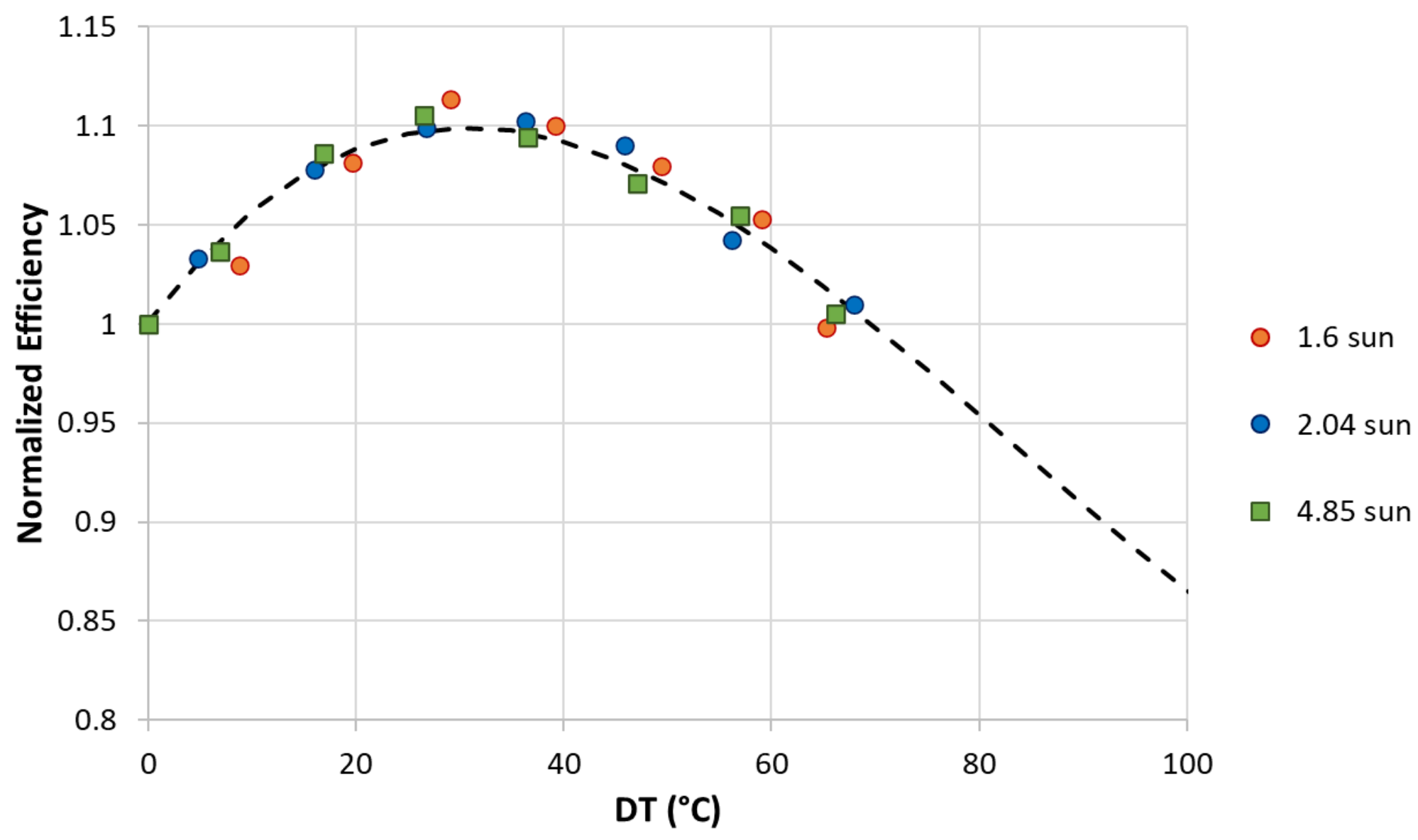

*Figure S5: PSC normalized efficiency versus temperature, at different optical concentrations. The black line is the fit implemented in the model reported in the main manuscript*

## S2 Bismuth telluride supplier's data

In Fig. S5 we report the electrical resistivity (ρ), the Seebeck coefficient (α), and the thermal conductivity (κ) of the p and n thermoelectric materials implemented in this work. The materials are 99.99% pure $Bi_2Te_3$, and $Sb_2Te_3$, hot pressed alloys. The reported data were acquired by the material supplier.

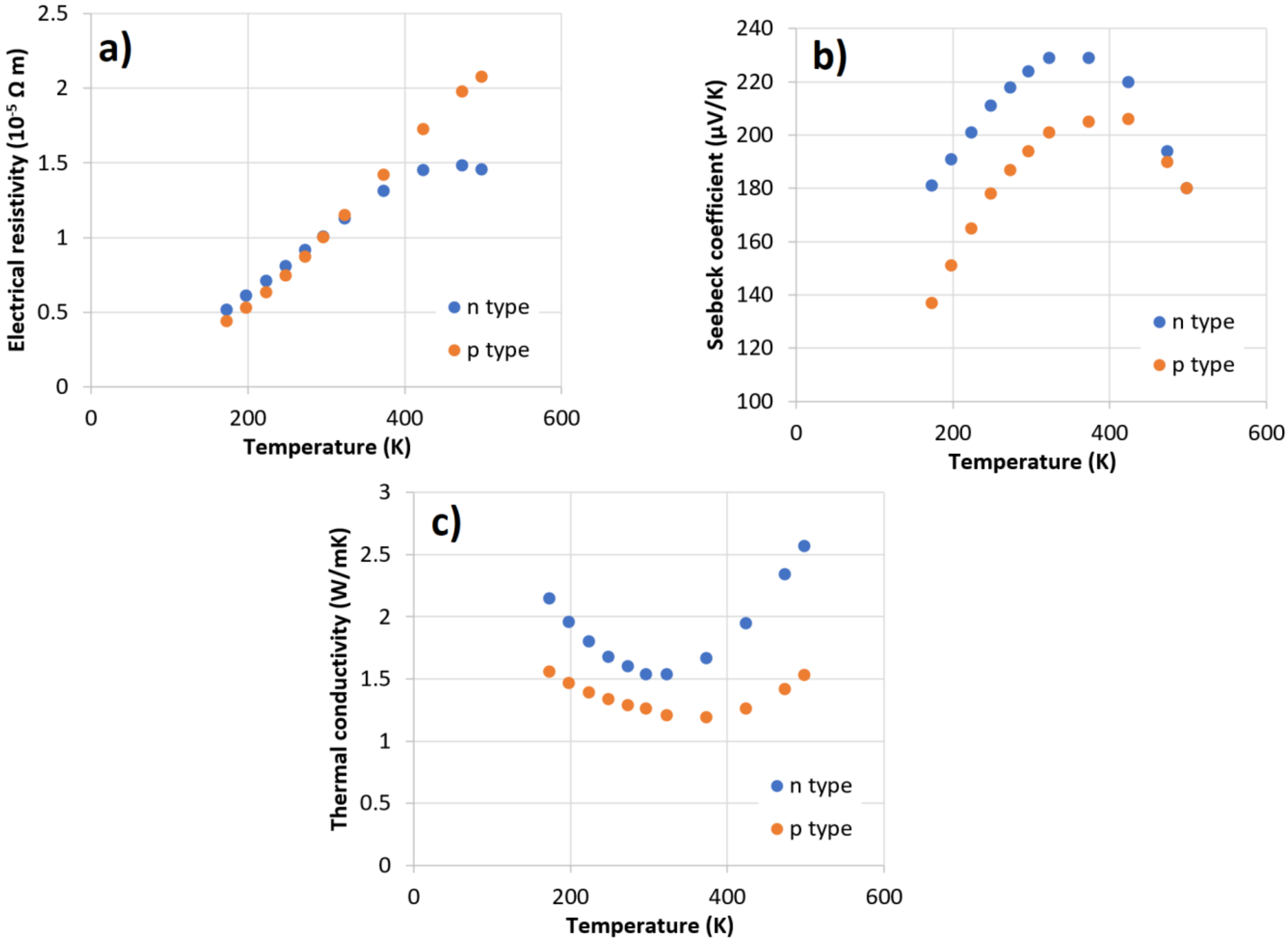

*Figure S5: Values of the electrical resistivity (a), Seebeck coefficient (b), and the thermal conductivity (c) of the p and n thermoelectric materials implemented in this work.*